\pdfoutput=1
\documentclass[12pt]{article}
\usepackage{amssymb}
\usepackage{subfig}
\usepackage{enumerate,booktabs}

\usepackage{graphicx,setspace,lscape,longtable,multirow}
\usepackage{natbib,epsfig,graphicx}
\usepackage{amsmath,amsthm,amssymb,color}
\bibpunct{(}{)}{;}{a}{,}{,}

\newtheorem{theorem}{Theorem}
\newtheorem{lemma}{Lemma}
\newtheorem{prop}{Proposition}
\newcommand{\csection}[1]
    {\begin{center}
        \stepcounter{section}
        {\bf\large\arabic{section}. #1}
    \end{center}
    \vspace{-0.15 cm}
}

\newcommand{\csubsection}[1]{
\vspace{-0.25 cm}
\begin{center}
\stepcounter{subsection}
{\it\arabic{section}.\arabic{subsection}. #1}
\end{center}
\vspace{-0.25 cm}
}

\def\beq{\begin{equation}}
\def\eeq{\end{equation}}
\def\beqr{\begin{eqnarray}}
\def\eeqr{\end{eqnarray}}
\def\beqrs{\begin{eqnarray*}}
\def\eeqrs{\end{eqnarray*}}
\def\bet{\begin{theorem}}
\def\eet{\end{theorem}}
\def\bel{\begin{lemma}}
\def\eel{\end{lemma}}
\def\bep{\begin{prop}}
\def\eep{\end{prop}}
\def\bg{\begin{figure}[tbph]\begin{center}}
\def\eg{\end{center}\end{figure}}

\def\bc{\begin{center}}
\def\ec{\end{center}}

\def\wh{\widehat}

\def\1{\mbox{\boldmath $1$}}

\def\mE{\boldsymbol{\mathcal E}}

\def\mR{\mathbb{R}}
\def\mS{\mathbb S}
\def\mL{\mathcal L}

\def\mY{\mathbb Y}
\def\mX{\mathbb X}

\def\sxx{\widehat\Sigma_{xx}}
\def\sxy{\widehat\Sigma_{xy}}

\def\Ht{\wh{\theta}}
\def\HD{\wh{\Delta}}

\def\gd{\wh\theta_{\operatorname{gd}}}

\def\ols{\wh\theta_{\operatorname{ols}}}

\def\cov{\mbox{cov}}

\def\argmin{\mbox{argmin}}

\numberwithin{equation}{section}

\begin{document}

\begin{center}
{\bf\Large Statistical Analysis of Fixed Mini-Batch Gradient \\ Descent Estimator}

Haobo Qi$^{1}$, Feifei Wang$^{2,3}$\footnote{The corresponding author. Email: feifei.wang@ruc.edu.cn}, and Hansheng Wang$^1$

{\it\small
$^1$ Guanghua School of Management, Peking University, Beijing, China;\\
$^2$ Center for Applied Statistics, Renmin University of China, Beijing, China;\\
$^3$ School of Statistics, Renmin University of China, Beijing, China.

}

\end{center}

\begin{singlespace}
\begin{abstract}
We study here a fixed mini-batch gradient decent (FMGD) algorithm to solve optimization problems with massive datasets. In FMGD, the whole sample is split into multiple non-overlapping partitions. Once the partitions are formed, they are then fixed throughout the rest of the algorithm. For convenience, we refer to the fixed partitions as fixed mini-batches. Then for each computation iteration, the gradients are sequentially calculated on each fixed mini-batch. Because the size of fixed mini-batches is typically much smaller than the whole sample size, it can be easily computed. This leads to much reduced computation cost for each computational iteration. It makes FMGD computationally efficient and practically more feasible. To demonstrate the theoretical properties of FMGD, we start with a linear regression model with a constant learning rate. We study its numerical convergence and statistical efficiency properties. We find that sufficiently small learning rates are necessarily required for both numerical convergence and statistical efficiency. Nevertheless, an extremely small learning rate might lead to painfully slow numerical convergence. To solve the problem, a diminishing learning rate scheduling strategy \citep{2019Understanding} can be used. This leads to the FMGD estimator with faster numerical convergence and better statistical efficiency. Finally, the FMGD algorithms with random shuffling and a general loss function are also studied.
\end{abstract}


\end{singlespace}

\newpage
\csection{INTRODUCTION}

Modern statistical analyses often encounter challenging optimization problems with massive datasets, ultrahigh dimensional features, and extremely complicated objective functions \citep{2015ImageNet,He2016Deep,2017Building,NEURIPS2021_019f8b94}. In these cases, the classical Newton-Raphson method can hardly be applied, because the second-order derivatives of the objective function (i.e., Hessian matrices) are extremely complicated. To solve the problem, various first-order optimization algorithms have been proposed and widely adopted in practice. Among them, the gradient descent (GD) type of algorithms are arguable the simplest and most widely used methods \citep{Cauchy1847}.

To illustrate the idea of the GD algorithm, let $(X_i, Y_i)$ be the observation collected from the $i$-th subject, where $Y_i\in\mathbb{R}^1$ is the response of interest and $X_i\in\mR^p$ is the associated $p$-dimensional predictor for $1\leq i \leq N$. To model the relationship between $X_i$ and $Y_i$, a parametric model with the parameter $\theta$ is defined. To estimate $\theta$, define a global loss function $\mL_{N}(\theta) =N^{-1}\sum^N_{i=1}\ell(X_i,Y_i; \theta)$, where $\ell(X_i,Y_i;\theta)$ is a loss function evaluated at the $i$-th sample. The GD algorithm would start with an initial estimate $\widehat{\theta}^{(0)}$ and then iteratively update according to the following formula,
\beq
\gd^{(t+1)}=\gd^{(t)}-\alpha\nabla\mL_N\big(\gd^{(t)}\big),\nonumber
\eeq
where $\gd^{(t)}$ is the estimator obtained in the $t$-th iteration, $\alpha$ is the learning rate, and $\nabla\mL_N(\theta)$ stands for the first-order derivative of $\mL_{N}(\theta)$ with respect to $\theta$.

For an optimization problem with $N$ samples and $p$ features, the computational complexity of GD algorithms for one single iteration is about $O(Np)$. It remains to be very expensive if both $N$ and $p$ are very large. Consider the classical deep learning model \emph{ResNet} \citep{He2016Deep} for example. A standard \emph{ResNet50} model contains more than 25 million parameters and has an extremely complicated nonlinear model structure. It achieves an excellent Top-1 accuracy (i.e., $83\%$) on the ImageNet 2012 classification dataset, which contains a total of 1,431,167 photos belonging to 1000 classes \citep{2015ImageNet}. Obviously, to estimate the {\it ResNet} model with the ImageNet dataset, the GD algorithm remains to be computationally very expensive. Then, how to further reduce the computation cost for massive datasets with ultrahigh dimensional features is worth of consideration.

One way to address this issue is to further reduce the sample size from $N$ to a much reduced number $n$. Then, the computation cost per iteration can be further reduced from $O(Np)$ to $O(np)$. This leads to the idea of fixed mini-batch gradient descent (FMGD) algorithm to be studied in this work. Specifically, assume the whole sample with size $N$ can be split into multiple non-overlapping partitions with each having $n$ samples. Once those partitions are formed, they should be fixed throughout the rest of the algorithm. For convenience, we refer to each fixed partition as a fixed mini-batch and denote its size by $n$. Subsequently, for each iteration, the current estimate should be updated using gradients calculated based on the $n$ samples in one fixed mini-batch. In theory, we might still need $n$ goes to infinity as $N$ goes to infinity. However, its diverging rate can be much slower than $N$. Consequently, the batch size $n$ can be comparatively much smaller than $N$. Then the computation cost of one single iteration can be largely reduced.

It is remarkable that the FMGD algorithm studied in this work is closely related to various stochastic mini-batch gradient descent (SMGD) algorithms, which have been extensively studied in the literature. Due to their outstanding computational performance, the SMGD algorithms (e.g., momentum, Adagrad, RMSprop, Adam) are also popularly applied in practice for sophisticated model learning \citep{2011Adagrad,2012RMSprop,2014Adam}. The FMGD algorithm proposed in this work is similar with the SMGD algorithms in the sense that, they both compute gradients on mini-batches. However, FMGD and SMGD are critically different from each other according to how the mini-batches are generated. Specifically, the mini-batches used by FMGD are fixed once they are formed. However, those of SMGD are randomly generated. In this regard, the SMGD methods can be further classified into two categories. The first category assumes that independent mini-batches can be directly sampled from the population distribution without limitation. This setting is very suitable for streaming data analysis \citep{Mou2020OnLS,An2020Yu,Chen2022StationaryBO}. The second category of the SMGD algorithms assume that the mini-batches are randomly sampled from the whole dataset by (for example) the method of simple random sampling with replacement \citep{OptinML}. By doing so, the whole sample is treated as if it were the population. Accordingly, the mini-batches become independent and identically distributed conditional on the whole sample. This makes the resulting theoretical analysis more convenient.

Different from various SMGD methods, the key feature of the FMGD algorithm is that the mini-batches used by FMGD are fixed and then repeatedly used once they are formed. As a consequence, the sequentially updated mini-batches and various statistics are not independent with each other, even if conditional on the whole sample. This makes the theoretical treatment of FMGD very different from that of SMGD in the literature. Accordingly, new techniques have to be developed. In this regard, we develop here a linear dynamic system approach, whose dynamic properties determine the numerical convergence rate of the FMGD algorithm. In the meanwhile, the stable solution of the linear dynamic systems determines the FMGD estimator and thus its asymptotic properties. It seems to us that, techniques of similar type have not been seen in the past literature \citep{Bottou2018Optimization,Bridging2020,First-order2020,Mou2020OnLS,An2020Yu,Chen2022StationaryBO}. With the help of these novel techniques, the numerical convergence properties of the FMGD algorithm can be well separated from the statistical convergence properties of the FMGD estimator. The former is driven by the number of the numerical iterations, while the latter is mainly determined by the whole sample size.

Specifically, we start with a linear regression model, and formulate the FMGD algorithm as a linear dynamic system. Then we investigate the conditions under which the iteratively updated FMGD estimators should converge to a stable solution. We refer to this stable solution as the FMGD estimator.  We then study the asymptotic properties of the resulting FMGD estimator under a constant learning rate. We find that, a standard FMGD estimator with a fixed learning rate might be statistically inefficient. To achieve both statistical efficiency and numerical convergence, a classical scheduling strategy with diminishing learning rate \citep{Numerical,2019Understanding} can be used. This leads to excellent finite sample performance for FMGD. Finally, we make two important extensions about FMGD. First, an FMGD algorithm with random shuffling is studied. Second, a more general loss function is investigated. Extensive simulation studies have been conducted to demonstrate the finite sample performance of the FMGD estimator. A number of deep learning related numerical examples are also presented.

To summarize, we aim to provide the following contributions to the existing literature. First, we develop here the FMGD algorithm with outstanding numerical convergence rate and statistical efficiency. Second, we develop a novel linear dynamic system framework to study the theoretical properties of the FMGD method. Numerically, we show that the FMGD algorithm converges much faster than its SMGD counterparts. Statistically, the resulting FMGD estimator enjoys the same asymptotic efficiency as the global estimator. The rest of this article is organized as follows. Section 2 introduces the FMGD algorithm under the linear regression setup. Section 3 discusses the theoretical properties of the FMGD estimator. Section 4 presents extensive numerical experiments to demonstrate the finite sample performance of the FMGD estimator. Section 5 concludes the paper with a brief discussion.

\csection{FIXED MINI-BATCH GRADIENT DESCENT}

\csubsection{A Linear Regression Setup}

Let $\mS=\{1,2,\cdots,N\}$ be the index set of the whole sample.
For each sample $i$, we collect a response variable of interest $Y_i\in\mathbb{R}^1$ and an associated $p$-dimensional predictor $X_i=(X_{i1},X_{i2},\cdots,X_{ip})^\top\in\mR^p$. We assume $N$ goes to infinity and $p$ is fixed. Different samples are assumed to be independently and identically distributed. Define the loss function evaluated at sample $i$ as $\ell(X_i,Y_i;\theta)$, where $\theta\in\mathbb{R}^{q}$ denotes the parameter. Then the global loss function can be constructed as $\mL_N(\theta)=N^{-1}\sum^N_{i=1} \ell(X_i,Y_i;\theta)$. The empirical risk minimizer $\Ht = \operatorname{\argmin}\mL_N(\theta)$ is a natural estimator of $\theta$.

We start with a linear regression model. Subsequently, the fruitful theoretical results obtained here can be extended to more general loss functions. Specifically, we start by assuming $Y_i=X_i^\top\theta+\varepsilon_i$, where $\theta = (\theta_1,\theta_2,\cdots,\theta_p)^\top\in\mR^p$ becomes
the regression coefficient vector and $\varepsilon_i$s ($1\leq i \leq N$) are mutually independent noises with mean 0 and variance $\sigma_\varepsilon^2$. Furthermore, we assume that all entries of $X_i$ are sub-gaussian in the sense that there exists a constant $K$ such that $P(|X_{ij}|>u)\leq \exp(1-u^2/K^2)$ for $1\leq i\leq N$ and $1\leq j\leq p$. Define the response vector as $\mY=(Y_1,Y_2,\cdots,Y_N)^\top\in\mR^N$
, the design matrix as $\mX=(X_1,X_2,\cdots,X_N)^\top\in\mR^{N\times p}$ and the noise vector as $\mE=(\varepsilon_1,\varepsilon_2,\cdots,\varepsilon_N)^\top\in\mR^N$. Then, the model can be re-written into a matrix form as
\begin{equation}
\label{eq:ols}
\mY=\mX\theta+\mE.
\end{equation}
Assume the loss function for the $i$-th sample is $\ell(X_i,Y_i;\theta) = (Y_i-X_i^\top\theta)^2/2$. Then, the global loss function becomes the least square loss function and the corresponding ordinary least squares (OLS) estimator can be obtained as $\ols=\argmin_\theta \mL_N(\theta)$ $= \sxx^{-1}\sxy$, where $\sxx=N^{-1}\sum X_iX_i^\top$ and $\sxy=N^{-1}\sum X_iY_i$. Standard asymptotic theory reveals that $\ols$
is $\sqrt{N}$-consistent and asymptotically normal. More specifically, we should have $\sqrt{N}(\ols-\theta)\rightarrow_d N(0,\sigma_\varepsilon^2\Sigma_{xx}^{-1})$,
where $\Sigma_{xx}=\cov(X_i)$.

\csubsection{A Standard Gradient Descent Algorithm}
\label{GD}

Next, we consider how to estimate $\ols$ by the method of gradient descent (GD). A standard GD algorithm is an iterative algorithm. It starts with an initial value $\widehat{\theta}^{(0)}$, which is often randomly generated or simply set to be zero. Then, the GD algorithm would iteratively update the current estimator to the next one, until the algorithm converges numerically. Let $\gd^{(t)}$ be the estimator obtained in the $t$-th iteration. Then, a GD algorithm should start with an initial estimator $\widehat{\theta}^{(0)}$ and then update $\gd^{(t)}$ in the $(t+1)$-th iteration as
\beq
\label{eq:gd}
\gd^{(t+1)}=\gd^{(t)}-\alpha\nabla\mL_N\big(\gd^{(t)}\big)=\HD_\alpha\gd^{(t)}+\alpha\sxy,
\eeq
where $\nabla\mL_N(\theta)=N^{-1}\sum^N_{i=1} \nabla \ell(X_i,Y_i; \theta)$ and $\nabla\ell(x,y; \theta)$ stands for the first-order derivative of $\ell(x,y; \theta)$ with respect to parameter $\theta$.
Here $\alpha>0$ is the learning rate and $\HD_\alpha=I-\alpha\sxx \in\mR^{p\times p}$ is the contraction matrix of the GD algorithm.  We refer to $\HD_\alpha$ as a contraction operator. According to standard linear system theory, the algorithm (\ref{eq:gd}) converges numerically if and only if the spectral radius of $\HD_\alpha$ is smaller than 1. Specifically, let $\lambda_j(A)$ be the $j$-th largest eigenvalue of an arbitrary matrix $A\in\mathbb{R}^{p\times p}$. For convenience, we also write $\lambda_1(A)=\lambda_{\max}$ and $\lambda_p(A)=\lambda_{\min}$. The spectral radius of $A$ can be defined as $\rho(A) = \max_{1\leq j\leq p}|\lambda_j(A)|$. To study the property of $\rho(\HD_\alpha)$, define the population counterpart of $\HD_\alpha$ as $\Delta_\alpha = I - \alpha\Sigma_{xx}$. One can verify that $\rho(\Delta_\alpha) = \max \big\{|1-\lambda_1(\Sigma_{xx})\alpha|,|1-\lambda_p(\Sigma_{xx})\alpha|\big\}$.

Ample amount of
empirical experiences suggest that, if $\alpha$ is set to be unnecessarily large, the
GD estimator might diverge and thus cannot converge numerically to any finite limit. In contrast, if $\alpha$ is set to be too small, then the GD algorithm might converge at a painfully slow speed. This interesting phenomenon can be theoretically explained by the following proposition.
	\bep
	\label{prop1}
	Let $(\HD_\alpha)^t$ be the $t$-th power of $\HD_\alpha$. Let $\|\cdot\|$ denote the $L_2$-norm for vectors and the induced $L_2$-norm for matrices. We then have: (1) $\gd^{(t)}=\big\{I-(\HD_\alpha)^t\big\}\ols
	+(\HD_\alpha)^t\widehat{\theta}^{(0)}$; and (2) for any $0< \alpha< 2/\lambda_1(\Sigma_{xx})$ and an arbitrary but sufficiently small $\eta>0$, there exists some positive constants $C$ and $\rho(\Delta_\alpha)-\eta<\rho_{\alpha}<1$ such that $P\Big(\|\gd^{(t)}-\ols\|\leq \rho_{\alpha}^t\|\widehat{\theta}^{(0)}-\ols\|\Big)\geq 1- 2\exp\Big(-CN\eta^2/\alpha^2\Big)$.
	\eep
	\noindent
More general conclusions in this regard have been obtained in the literature. See for examples, \cite{2009Robust}, \cite{2012sgd}, \cite{2016SGD}, \cite{2017AoS}, and \cite{2019Understanding}. For the sake of theoretical completeness, we provide the detailed proof of Proposition \ref{prop1} in Appendix A.1. By Proposition \ref{prop1}, we find that the GD estimator $\gd^{(t)}$ enjoys an analytically explicit solution. It is a convex combination of the OLS estimator and the initial value. When the learning rate satisfies $0< \alpha< 2/\lambda_1(\Sigma_{xx})$, we have $\rho(\Delta_\alpha)<1$. Consequently, for an arbitrarily but sufficiently small $\eta>0$, there exists a constant $\rho(\Delta_\alpha)-\eta<\rho_{\alpha}<1$ such that $\|\gd^{(t)}-\ols\|\leq \rho_{\alpha}^t\|\widehat{\theta}^{(0)}-\ols\|$ holds with probability no smaller than  $1-2\exp(-CN\eta^2/\alpha^2)$. Note that this convergence rate is mainly due to the fact $\|\gd^{(t)}-\ols\|\leq \rho_{\alpha}^t\|\widehat{\theta}^{(0)}-\ols\|$. In other words, the numerical error of the $t$-th step estimator $\gd^{(t)}$ is linearly bounded by that of the previous step with large probability. Following \cite{1963Polyak}, \cite{Introductory2004}, \cite{2016Linear}, and \cite{2020Linear}, we refer to this interesting property as a linear convergence property.

\csubsection{The Fixed Mini-Batch Gradient Descent Algorithm}
\label{minibatchGD}

To implement the FMGD algorithm, the whole sample should be randomly divided into a total of $M$ non-overlapping mini-batches in each epoch, where $M$ might be pre-specified by the user. Denote $\{\mS^{(t,m)}\}^M_{m=1}$ as the mini-batch index sets in the $t$-th epoch. We should have $\mS=\bigcup_m\mS^{(t,m)}$ and $\mS^{(t,m_1)}\bigcap \mS^{(t,m_2)} = \emptyset$ for any $t\geq 1$ and $m_1\neq m_2$. For convenience, assume $N$ and $M$ are particularly designed so that $n=N/M$ is an integer. We then assume all mini-batches have the same sample size as $|\mS^{(t,m)}|=n$ for $t\geq 1$ and $1 \leq m \leq M$. For the FMGD algorithm, once the mini-batch $\mS^{(t,m)}$ is formed, it should be fixed throughout the rest of the algorithm. Consequently, we should have $\mS^{(t,m)}=\mS^{(m)}$ for some $\mS^{(m)} \subset \mS$ and for any $t\geq 1$. Obviously, we should have $\bigcup_m\mS^{(m)}=\mS$ and $\mS^{(m_1)}\bigcap \mS^{(m_2)} = \emptyset$ for any $m_1\neq m_2$.

Recall $\Ht^{(0)}$ is the initial estimator. Practically, $\Ht^{(0)}$ is often randomly generated or simply set to be zero. Let $\Ht^{(t,m)}$ be the FMGD estimator obtained in the $t$-th epoch on the $m$-th mini-batch. We then have the following updating formula for the FMGD algorithm as
\beq
\begin{split}
\label{eq:3}
\Ht^{(t,1)}&=\Ht^{(t-1,M)}-\alpha\nabla\mL_n^{(1)}\Big(\Ht^{(t-1,M)}\Big),\\
\Ht^{(t,m)}&=\Ht^{(t,m-1)}-\alpha\nabla\mL_n^{(m)}\Big(\Ht^{(t,m-1)}\Big) \mbox{ for } 2\leq m \leq M,
\end{split}
\eeq
where $\nabla\mL_n^{(m)}(\theta)=n^{-1}\sum_{i\in\mS^{(m)}}\nabla\ell(X_i,Y_i;\theta)$ is the gradient computed in the $t$-th epoch on the $m$-th mini-batch, and $\Ht^{(0,M)}=\Ht^{(0)}$. Under the linear regression setting, one can easily verify that $\nabla\mL_n^{(m)}(\theta)=\sxx^{(m)}\theta-\sxy^{(m)}$ with
$\sxx^{(m)}=n^{-1}\sum_{i\in\mS^{(m)}} X_iX_i^\top$ and $\sxy^{(m)}=n^{-1}\sum_{i\in\mS^{(m)}}X_i Y_i$.
Then, \eqref{eq:3} can be re-written as
\beq
\begin{split}
\label{eq:2}
\Ht^{(t,1)}&=\HD_\alpha^{(1)}\Ht^{(t-1,M)}+\alpha\sxy^{(1)},\\
\Ht^{(t,m)}&=\HD_\alpha^{(m)}\Ht^{(t,m-1)}+\alpha\sxy^{(m)} \mbox{ for } 2\leq m \leq M,
\end{split}
\eeq
where $\HD_\alpha^{(m)}=I-\alpha\sxx^{(m)}$ is the contraction operator generated by the $m$-th mini-batch in the $t$-th epoch. For convenience, we refer to \eqref{eq:2} as an FMGD algorithm for the linear regression model.

\csubsection{Comparing FMGD with SMGD}

As we mentioned before, the FMGD algorithm is similar with the SMGD algorithms extensively studied in the literature. To compare the differences between FMGD and SMGD, we rewrite the FMGD algorithm \eqref{eq:3} and \eqref{eq:2} in a more general theoretical framework \citep{Mou2020OnLS, An2020Yu} as
\begin{eqnarray}
\label{sgd}
\Ht^{(t,1)}=\Ht^{(t-1,M)}-\alpha \left\{\nabla\mL\Big(\wh\theta^{(t-1,M)}\Big) + \xi_{t,1}\Big(\wh\theta^{(t-1,M)}\Big)\right\},\ \ \ \ \ \ \ \ \ \notag \\
\Ht^{(t,m)}=\Ht^{(t,m-1)}-\alpha\left\{\nabla\mL\Big(\wh\theta^{(t,m-1)}\Big) + \xi_{t,m}\Big(\wh\theta^{(t,m-1)}\Big)\right\} \mbox{ for } 2\leq m \leq M.
\end{eqnarray}
Here $\nabla\mL(\theta) = E\{\nabla\ell(X_i,Y_i; \theta)\}=\Sigma_{xx}\theta-\Sigma_{xy}$ with $\Sigma_{xy}=E(X_iY_i)$ is the population gradient. Moreover, $\xi_{t,m}(\theta)=\nabla\mL_n^{(m)}(\theta)-\nabla\mL(\theta)=(\sxx^{(m)}-\Sigma_{xx})\theta-(\sxy^{(m)}-\Sigma_{xy})$ can be viewed as a zero-mean noise added to the population gradient. This makes the computed gradient \emph{stochastic}. Note that the theoretical framework \eqref{sgd} has been popularly used in the literature to study the theoretical properties of various SMGD algorithms \citep{Bridging2020, Mou2020OnLS, An2020Yu,Chen2022StationaryBO}. It seems that our FMGD method can also be nicely covered by this elegant theoretical framework. By this framework, the difference between FMGD and SMGD can be better illustrated as follows.

By  \eqref{sgd} we find that, FMGD and SMGD are indeed very similar with each other. However, these exists one critical difference. That is how the random noise term $\xi_{t,m}(\theta)$ is generated. For most SMGD algorithms studied in the literature, $\xi_{t,m}(\theta)$ is always assumed to be independently generated for different $t$ and $m$. The independence could be marginal independence due to (for example) the unlimited streaming data, or conditional independence due to (for example) subsampling on the whole dataset. In this case, the subscript $m$ for identifying different mini-batches within the same epoch $t$ becomes unnecessary. This explains why in the existing SMGD literature, the subscript $m$ was seldom used. For most cases, one single subscript $t$ for identifying the number of numerical iterations is already sufficient. In contrast, for the FMGD algorithm, we have fixed partitions. The fixed partitions are repeatedly used throughout the whole algorithm. In other words, we always have $\mS^{(t_1,m)}=\mS^{(t_2,m)}=\mS^{(m)}$ for any $t_1\neq t_2$ but a fixed $m$. Obviously, we should have $\bigcup_m\mS^{(m)}=\mS$ and $\mS^{(m_1)}\bigcap \mS^{(m_2)} = \emptyset$ for any $m_1\neq m_2$. Accordingly, the random noise terms used in \eqref{sgd} can never be independent with each other, neither marginally nor conditionally. This is because $\xi_{t_1,m}(\theta)=\xi_{t_2,m}(\theta)$ for any $t_1\neq t_2$ since they are both computed on the same mini-batch $\mS^{(m)}$.

To further illustrate the idea, we compare the two algorithms (i.e., FMGD and SMGD) under a linear regression model setup with the least squared loss function. Let $\Ht_{\text{fmgd}}^{(t,M)}$ be the FMGD estimator obtained in the $t$-th epoch on the $m$-th mini-batch and $\Ht_{\text{smgd}}^{(t,M)}$ be the SMGD estimator obtained after $tM$-th mini-batch updates. Then by equation (\ref{sgd}), we know that both the FMGD and SMGD methods should share a similar updating formula as follows
\beqr\label{update}
\Ht^{(t,m)}-\ols = \HD_\alpha^{(t,m)}\left(\Ht^{(t,m-1)}-\ols\right) + \alpha\left(\sxy^{(t,m)} - \sxx^{(t, m)}\ols\right),
\eeqr
where $\Ht^{(t,m)}$ can be $\Ht_{\text{fmgd}}^{(t,M)}$ or $\Ht_{\text{smgd}}^{(t,M)}$. By iteratively applying equation (\ref{update}) for a total of $M$ times (i.e., a single epoch), one can obtain that $\Ht^{(t,M)}-\ols = Q_1 + Q_2$, where
\beqrs
&&Q_1 = \left(\prod^M_{m=1}\HD^{(t,m)}_\alpha\right)\Big(\Ht^{(t-1,M)}-\ols\Big),\\
&&Q_2 = \alpha \sum^M_{m=1}\left(\prod^M_{s=m+1}\HD^{(t,m)}_\alpha\right)\bigg(\sxy^{(t,m)} - \sxx^{(t,m)}\ols\bigg).
\eeqrs
For both FMGD and SMGD, $Q_1$ represents the contracted distance between historical estimator and the OLS estimator, while $Q_2$ represents the accumulated computational error due to the mini-batch gradients. It is remarkable that for the FMGD algorithm, $\sxx^{(t,m)}$s and $\sxy^{(t,m)}$s are statistics calculated on non-overlapping subsets of the whole sample. Therefore, they satisfy an interesting analytical relationship as $\sxx = M^{-1}\sum^M_{m=1}\sxx^{(t,m)}$ and $\sxy = M^{-1}\sum^M_{m=1}\sxy^{(t,m)}$. Consequently, we have $\sum^M_{m=1}(\sxy^{(t,m)} - \sxx^{(t, m)}\ols) = 0$ for every $t>0$, since $\ols = \sxx^{-1}\sxy$. Then the $Q_2$ term for the FMGD algorithm reduces to
\beqrs
Q_2 = \alpha \sum^M_{m=1}\left(\prod^M_{s=m+1}\HD^{(t,m)}_\alpha-I\right)\bigg(\sxy^{(t,m)} - \sxx^{(t,m)}\ols\bigg)
\eeqrs
with $\|\prod^M_{s=m+1}\HD^{(t,m)}_\alpha-I\| = O_p(\alpha)$. In the meanwhile, under appropriate regularity conditions, we can verify that $\|\sxy^{(t,m)} - \sxx^{(t,m)}\ols\| = O_p(n^{-1/2})$ is bounded in probability for all $1\leq m\leq M$. As a result, the $Q_2$ term in FMGD becomes $O_p(\alpha^2n^{-1/2})$, which is a term of very tiny size for small learning rate $\alpha$. In contrast, for the SMGD algorithm, both $\sxx^{(t,m)}$s and $\sxy^{(t,m)}$s are statistics calculated on randomly sampled subsets of the whole sample with replacement. Consequently, we have $\sum^M_{m=1}(\sxy^{(t,m)} - \sxx^{(t, m)}\ols) =O_p(n^{-1/2})$ for every $t$. Then the $Q_2$ term remains to be $O_p(\alpha n^{-1/2})$, instead of $O_p(\alpha^2 n^{-1/2})$ in the FMGD algorithm. This suggests that the accumulated computational error by FMGD in every epoch iteration is smaller than the counterpart of the SMGD algorithm. This is mainly because in each epoch of the FMGD method, every sample point in the whole sample is guaranteed to be fully utilized. However, this nice property cannot be achieved by SMGD due to its stochastic nature for mini-batch sampling. This makes the numerical convergence performance of SMGD less efficient.

\csection{THEORETICAL PROPERTIES}

\csubsection{The Stable Solution}

To study the theoretical properties of FMGD, we start with a linear regression model with a fixed learning rate. Comparing \eqref{eq:gd} with \eqref{eq:2}, we find that only one contraction operator $\HD_\alpha$ is involved in the GD estimator, while a total of $M$ contraction operators (i.e., $\HD_\alpha^{(m)}$) are involved in the FMGD algorithm. As a result, the FMGD algorithm is considerably more sophisticated. Moreover, it is natural to query whether a numerical limit of the FMGD sequence does exist as $t\rightarrow\infty$. If such a limit exists, we refer to it as the FMGD estimator. Before we formally address this important problem, we temporally assume that, there exists a limit $\Ht^{(m)}$ such that $\Ht^{(t,m)}\rightarrow \Ht^{(m)}$ as $t\rightarrow\infty$ for $1\leq m\leq M$. It following then $\Ht^{(m)}$ should be the stable solution of the following interesting linear dynamic system:
\beq
\begin{split}
	\label{eq:4}
	\Ht^{(1)}&=\HD_\alpha^{(1)}\Ht^{(M)}+\alpha\sxy^{(1)},\\
	\Ht^{(m)}&=\HD_\alpha^{(m)}\Ht^{(m-1)}+\alpha\sxy^{(m)} \mbox{ for } 2\leq m \leq M.
\end{split}
\eeq
It is remarkable that the contraction operator $\HD_\alpha^{(m)}$ does not depend on $t$. Write the linear dynamic system \eqref{eq:4} into a matrix form as $\wh\Omega \wh\theta^*=\alpha\sxy^*$, where
$\wh\theta^* = (\Ht^{(1)\top},\Ht^{(2)\top},\cdots,\Ht^{(M)\top})^\top\in\mR^{(Mp)}$ is the vectorized FMGD estimator,
$\sxy^*=(\sxy^{(1)\top},\sxy^{(2)\top},$ $\cdots,\sxy^{(M)\top})^\top\in\mR^{(Mp)}$, and $\wh\Omega$ is a $(Mp)\times (Mp)$ matrix given by
\[\wh\Omega=\left[\begin{array}{cccccc}
I&0&0&\cdots&0&-\HD_\alpha^{(1)}\\
-\HD_\alpha^{(2)}&I&0&\cdots&0&0\\
0&-\HD_\alpha^{(3)}&I&\cdots&0&0\\
\cdots&\cdots&\cdots&\cdots&\cdots&\cdots\\
0&0&0&\cdots&-\HD_\alpha^{(M)}&I
\end{array}
\right].\]
If the matrix $\wh\Omega$ is invertible, we should immediately have $\wh\theta^*=\alpha{\wh\Omega}^{-1}\sxy^*$. It is then of great interest to study two important problems. First, under what conditions the matrix $\wh\Omega$ is invertible? Once $\wh\Omega$ is invertible, the stable solution to the dynamic system \eqref{eq:4} can be uniquely determined. Second, what is the analytical solution for ${\wh\Omega}^{-1}$? Regarding the first problem, we give the following proposition.
\bep
\label{prop3}
 Assume $0<\alpha <2/\lambda_1(\Sigma_{\text{xx}})$. Then there exists a positive constant $C$ such that for any $0<\eta< \min\{\alpha \lambda_p(\Sigma_{\text{xx}}), 2-\alpha\lambda_1(\Sigma_{\text{xx}})\}$, $\wh\Omega$ is invertible with probability no less than $1 - 2M\exp\left\{-CN\eta^2/(\alpha^2M^2)\right\}$.
\eep
\noindent
The detailed proof is given in Appendix A.2. By Proposition \ref{prop3}, we find that the dynamic system \eqref{eq:4} should have a unique stable solution with large probability. However, whether this solution can be approached by the FMGD algorithm is another issue. An intuitive condition is that the learning rate $\alpha$ cannot be too large, otherwise the FMGD algorithm might not converge. If the stable solution can indeed be captured by the FMGD algorithm (in the sense that $\Ht^{(t,m)}\rightarrow \Ht^{(m)}$ as $t\rightarrow\infty$), we then refer to this stable solution as the FMGD estimator.
\csubsection{Numerical Convergence and Asymptotic Distribution}

In the previous subsection, we have studied the conditions for the existence of the stable solution. However, this does not necessarily imply the existence of the FMGD estimator. We refer to a stable solution $\Ht^{*}$ as an FMGD estimator only if it can be computed by the FMGD algorithm. Otherwise, it is just a stable solution to the dynamic system \eqref{eq:4}, but not a practically computable estimator. To find the answer, define the iteratively updated FMGD estimators at the $t$-th epoch as $\wh\theta^{*(t)}=(\Ht^{(t,1)\top},\cdots,\Ht^{(t,M)\top})^\top$. Next, we subtract both sides of \eqref{eq:4} from those of \eqref{eq:2}. We then obtain $\Ht^{(t+1,m)} - \Ht^{(m)} = \wh{C}^{(m)} \Big(\Ht^{(t,m)} - \Ht^{(m)}\Big)$ for $1\leq m\leq M$, where $\wh{C}^{(1)} =\HD_\alpha^{(1)}\HD_\alpha^{(M)}\cdots\HD_\alpha^{(2)}$, $\wh{C}^{(m)} =\HD_\alpha^{(m)}\cdots\HD_\alpha^{(1)}\HD_\alpha^{(M)}\cdots\HD_\alpha^{(m+1)}$ for $2 \leq m \leq M-1$, and $\wh{C}^{(M)} =\HD_\alpha^{(M)}\cdots\HD_\alpha^{(1)}$. Recall that $\rho(A)$ denotes the spectral radius of an arbitrary $p\times p$ real value matrix $A$. Then, the iteratively updated FMGD estimator $\Ht^{(t,m)}$ converges linearly towards the FMGD estimator $\Ht^{(m)}$, as long as  the spectral radius $\rho(\wh{C}^{(m)})<1$ for $1\leq m \leq M$. We summarize this interesting finding in the following theorem.

\bet
\label{Th1}
{\sc (Numerical Convergence)} Assume $0<\alpha<2/\lambda_{1}(\Sigma_{xx})$. Let $\Ht^{(0,m)}$ be the initial point in the $m$-th mini-batch with $\Ht^{(0,1)}=\Ht^{(0)}$ and $\Ht^{(0,m)}=\Ht^{(1,m-1)}$ for $2\leq m \leq M$. Then for an arbitrarily but sufficiently small $\eta>0$, there exist some positive constants $C$ 
and  $\{\rho(\Delta_\alpha)\}^M-\eta<\rho_{\alpha,M}<1$, such that
\begin{equation}
P\left(\Big\|\Ht^{(t,m)}-\Ht^{(m)}\Big\|\leq \rho_{\alpha,M}^{t-1}\Big\|\Ht^{(0,m)}-\Ht^{(m)}\Big\|\right)
\geq 1- 4M\exp\left(-\frac{CN\eta^2}{\alpha^2M^2}\right).
\nonumber
\end{equation}
\eet\noindent
The detailed proof of Theorem \ref{Th1} is given in Appendix A.3. Theorem \ref{Th1} suggests that, for any learning rate satisfying $0< \alpha< 2/\lambda_1(\Sigma_{xx})$ and an arbitrary but sufficiently small $\eta>0$, there exists a constant $\rho_{\alpha,M}<1$, such that $\|\Ht^{(t,m)}-\Ht^{(m)}\|\leq \rho_{\alpha,M}^{t-1}\|\Ht^{(0,m)}-\Ht^{(m)}\|$ holds with probability no less than  $1- 4M\exp(-CN\eta^2/\alpha^2M^2)$. Therefore, the FMGD algorithm can converge linearly to its stable solution. Here by ``converge linearly", we mean that the numerical error of the $t$-th step FMGD estimator is linearly bounded by that of the previous step \citep{1963Polyak,2009Robust}. In addition we find the sufficient conditions for the numerical convergence of the FMGD algorithm is more restrictive than the conditions for the existence of the stable solution. Last, we focus on the influence of learning rate $\alpha$ on the numerical convergence rate of FMGD. Specifically, the convergence factor is given by $\rho_{\alpha,M}=\rho(\wh{C}^{(M)})$, where $\wh{C}^{(M)} =\HD_\alpha^{(M)}\cdots\HD_\alpha^{(1)}$. Its population counterpart is given by $\{\rho(\Delta_\alpha)\}^M = \max\{|1-\alpha\lambda_1(\Sigma_{xx})|^M, |1-\alpha\lambda_p(\Sigma_{xx})|^M\}$. Therefore, as $\alpha \to 0$, we should have $\{\rho(\Delta_\alpha)\}^M \to 1$. This forces the convergence factor $\rho_{\alpha,M}\rightarrow 1$ in probability and thus leads to very slow numerical convergence rate for $\Ht^{(t,m)}$. This suggests that small $\alpha$ value leads to slow numerical convergence. We next study the statistical properties of the FMGD estimator, which is summarized in the following theorem.

\bet
\label{Th2}
{\sc (Asymptotic Normality)} Suppose the assumptions in Theorem \ref{Th1} hold. Then conditional on the whole sample, we have $\sqrt{N}\{v_m(\alpha)\}^{-1/2}\{\wh\theta^{(m)} - \wh{\mu}_m(\alpha)\}\rightarrow_d N(0,\sigma_\varepsilon^2I_p)$ as $N \rightarrow \infty$, where
\begin{equation}
\wh{\mu}_m(\alpha)=\theta + O_p\left(\alpha^2 n^{-3/2}\right) \mbox{~~and~~~}
v_m(\alpha)= \Sigma^{-1}_{xx} + \frac{\alpha^2(M^2-1)}{12}\Sigma_{xx} + o(\alpha^2).\nonumber
\end{equation}
\eet
\noindent
The detailed proof is given in Appendix A.4. Theorem \ref{Th2} suggests that, with an appropriately specified constant learning rate $\alpha$, the FMGD estimator $\wh\theta^{(m)}$ might still converge at the $\sqrt{N}$ speed. However, it is biased and its variance is larger than that of the whole sample OLS estimator by an appropriate amount of $\alpha^2\sigma_\varepsilon^2(M^2-1)\Sigma_{xx} /12$. Furthermore, a smaller learning rate should lead to smaller bias and smaller variance. The variance inflation effect disappears as $\alpha \rightarrow 0$.

\csubsection{Learning Rate Scheduling}

Previous analyses suggest that the FMGD estimator with a constant learning rate can hardly achieve the same asymptotic efficiency as the whole sample OLS estimator, unless an extremely small learning rate is used. This is because the statistical efficiency requires a sufficiently small learning rate $\alpha$. However, an extremely small learning rate would lead to painfully slow numerical convergence. A more common practice is to use relatively large learning rates in the early stage of the FMGD algorithm. By doing so, the numerical convergence speed can be much improved. In the meanwhile, much smaller learning rates should be used in the late stage of the FMGD algorithm. By doing so, better statistical efficiency can be obtained. Then how to appropriately schedule the learning rates for the FMGD algorithm becomes the key issue.

To address this problem, we follow the classical literature and consider a scheduling strategy with diminishing learning rate \citep{Numerical,2019Understanding}. Specifically, we allow different epochs to have different learning rates. Let $\alpha_t$ be the learning rate used for the $t$-th epoch iteration. By diminishing the learning rate, we mean $\alpha_t$ should monotonically converges to 0 as $t \to \infty$ with an appropriate rate. Intuitively, the diminishing rate cannot be too fast, otherwise the algorithm might not converge numerically. In the meanwhile, it cannot be too slow, otherwise the resulting estimator should be statistically inefficient. Then what type of diminishing rate can satisfy both requirements is the key problem. We then have the following theorem to address this important problem.
\bet
\label{Th3}{\sc (Asymptotic Convergence)}
Assume that the whole data $\{(X_i,Y_i):1\leq i \leq N\}$ are given and fixed. Assume that there exist two positive constants $0<\lambda_{\min}\leq \lambda_{\max}<+\infty$ such that $\lambda_{\min}\leq \lambda_{p}(\sxx^{(m)})\leq \lambda_{1}(\sxx^{(m)}) \leq \lambda_{\max}$ for all $1\leq m \leq M$. Let $\Ht^{(t,M)}$ be the FMGD estimator obtained in the $t$-th epoch on the $M$-th mini-batch. Define $\nabla{\mL}_{\max} = \max_{m}\|\sxy^{(m)} - \sxx^{(m)}\ols\|$. Assume that $\alpha_t$ is the learning rate used in the $t$-th epoch, then we have
\beqr\label{eq:32}
\Big\|\Ht^{(t,M)} - \ols\Big\|\leq \frac{\big\|\Ht^{(0)} - \ols\big\|}{\exp\left\{M\lambda_{\min}\left(\sum^t_{k=1}\alpha_k\right)\right\}} + M\nabla{\mL}_{\max} \left(\frac{\lambda_{\max}}{\lambda_{\min}}\right)\sum^t_{k=1} \frac{\alpha^{2}_k}{\left(\sum^{t}_{s=k+1}\alpha_s\right)}.
\eeqr
FUrthermore, if the learning rate sequence satisfies three conditions as: (1) $\sum^\infty_{t=1}\alpha_t = \infty$, (2) $\sum^\infty_{t=1}\alpha^2_t < \infty$, and (3) $0<\alpha_1<\lambda^{-1}_{\max}$. Then we have $\|\Ht^{(t,M)} - \ols\| \to 0$ as $t\to \infty$.
\eet
The detailed proof of Theorem \ref{Th3} is given in Appendix A.6. By Theorem \ref{Th3}, we can obtain the following conclusions. The first term in the right hand side of \eqref{eq:32} explains how the initial value $\widehat{\theta}^{(0)}$ might affect the numerical convergence speed. In this regard, the learning rate sequence $\alpha_t$ should not diminish too fast in the sense we should have $\sum^\infty_{t=1}\alpha_t = \infty$. Otherwise the numerical error due to the initial value might not disappear. For this term, a larger number of mini-batches (i.e., $M$) leads to a larger number of numerical updates within one single epoch and thus a smaller numerical error. The second term in the right hand side of \eqref{eq:32} contains many factors. First, it reflects the fact that a too large number of mini-batches (i.e., $M$) is not necessarily good for the overall convergence. This is mainly because a too large mini-batch number (i.e., $M$) makes the sample size per mini-batch (i.e., $n=N/M$) too small. Second, it also reflects the effect due to the learning rate scheduling strategy. Under the assumption $\sum^\infty_{t=1}\alpha_t = \infty$, we have $1/\left(\sum^{t}_{s=k+1}\alpha_s\right)$ shrinks to 0 as $t\rightarrow \infty$ for any fixed $k$. Nevertheless, we have an infinite number of those terms, whose collective effect might not diminish, unless appropriate control can be provided. Fortunately, a nice control can be provided by $\alpha_t^2$ under the condition $\sum^{\infty}_{t=1} \alpha_t^2 < \infty$. Consequently we can apply dominate convergence theorem to \eqref{eq:32}. This makes the second term in the right hand side of \eqref{eq:32} shrink to 0. Lastly, we find that the spectrum of $\sxx^{(m)}$s also play an important role here. It suggests that the smaller the conditional number (i.e., $\lambda_{\max}/\lambda_{\min}$) is, the better the numerical convergence rate should be.

Next we discuss several commonly used learning rate decaying strategies. We start with the polynomial decay strategy as $\alpha_t=c_{\alpha}t^{-\gamma}$ for some positive constants $c_{\alpha}>0$ and $\gamma>0$. Then $0.5 < \gamma \leq 1$ should satisfy the theorem conditions. In contrast, the theorem conditions should be violated if $0<\gamma\leq 0.5$ or $\gamma>1$. Another one is the exponential decaying strategy $\alpha_t=c_\alpha\gamma^{-t/b}$, where $c_\alpha>0$ is a pre-defined initial learning rate, $0<\gamma<1$ is the decay rate, and $b$ is the decay step. The exponential decaying strategy dose not satisfy the condition $\sum^\infty_{t=1}\alpha_t = \infty$ and thus could lead to early stop. Finally we focus on the stage-wise step decaying strategy. In this strategy, the whole training procedure is cut into several stages according to the decay step $b$. Then a stage-wise constant learning rate is used for each stage. Depending on how the stage-wise constant learning rates are specified, the technical conditions in Theorem \ref{Th3} might be satisfied or violated. For example, if we set $\alpha_t=c_\alpha/t$ with $t$ representing the number of stages for some initial value $c_\alpha$, then the theorem conditions are satisfied. However, if we set $\alpha_t=c_{\alpha}\gamma^{k}$ with $c_{\alpha}>0$ and some $0<\gamma<1$, then the theorem conditions are violated.

\csubsection{Two Useful Extensions}

In this subsection, we make two useful extensions about FMGD. First, we develop a more flexible FMGD algorithm, which allows random shuffling about the whole sample between two consecutive epochs. Second, a more general loss function (e.g., the negative log-likelihood function) is studied. We start with the shuffled FMGD method (or sFMGD for short). In the original FMGD method discussed above, once the mini-batches are partitioned, they are then fixed throughout the whole algorithm. In the meanwhile, many practically implemented softwares (e.g., TensorFlow, PyTorch) allow the whole sample to be randomly shuffled between two consecutive epochs. Accordingly, we should have $\bigcup_m\mS^{(t,m)}=\mS$ and $\mS^{(t,m_1)}\bigcap \mS^{(t,m_2)}= \emptyset$ for any $m_1\neq m_2$ and a common $t$. However, we should have $\mS^{(t_1,m)}\neq \mS^{(t_2,m)}$ for $t_1 \neq t_2$ because the partitions are randomly shuffled for every $t$. Denote the shuffled FMGD estimator by $\widehat\theta^{(t,M)}_{\text{sf}}$. Then the updating formula for a linear regression model becomes
\beq
\begin{split}
\label{eq:sfmgd}
\Ht^{(t,1)}_{\text{sf}}&=\HD_\alpha^{(t,1)}\Ht^{(t-1,M)}_{\text{sf}}+\alpha\sxy^{(t,1)},\\
\Ht^{(t,m)}_{\text{sf}}&=\HD_\alpha^{(t,m)}\Ht^{(t,m-1)}_{\text{sf}}\alpha\sxy^{(t,m)} \mbox{ for } 2\leq m \leq M,
\end{split}
\eeq
where $\sxx^{(t,m)}=n^{-1}\sum_{i\in\mS^{(t,m)}} X_iX_i^\top$, $\sxy^{(t,m)}=n^{-1}\sum_{i\in\mS^{(t,m)}}X_i Y_i$, and $\HD_\alpha^{(t,m)}=I-\alpha\sxx^{(t,m)}$. Comparing \eqref{eq:sfmgd} and \eqref{eq:2}, we find the only difference between FMGD and sFMGD lies in the way that $\mS^{(t,m)}$ is generated. For sFMGD, the mini-batches $\mS^{(t,m)}$ with $1\leq m \leq M$ used in the $t$-th epoch are still generated under the constraints: (1) $\bigcup_m\mS^{(t,m)}=\mS$ and (2) $\mS^{(t,m_1)}\bigcap \mS^{(t,m_2)}= \emptyset$ for any $m_1\neq m_2$. Therefore different $\mS^{(t,m)}$s are not conditionally independent with each epoch. Consequently, sFMGD remains to be different from the SMGD algorithms studied in literature in terms of how the mini-batches are generated.

Next, we investigate the theoretical properties of the sFMGD algorithm. As the sFMGD algorithm randomly shuffles the mini-batches for every epoch, the stable solution no longer exists. We thus have to take a slightly different approach (i.e., error bound analysis) to understand its theoretical properties. We focus on the last mini-batch estimator in each epoch (i.e., $\Ht^{(t,M)}_{\text{sf}}$), because this is the estimator utilizing the whole sample information in a single epoch. Then we have the following theorem for the numerical error bound.

\bet
\label{Th4}{\sc (Numerical Error Bound)}
Assume that the whole data $\{(X_i,Y_i):1\leq i \leq N\}$ are given and fixed. Further assume that there exist two positive constants $0<\lambda_{\min}\leq \lambda_{\max}<+\infty$ such that $\lambda_{\min}\leq \lambda_{p}(\sxx^{(t,m)})\leq \lambda_{1}(\sxx^{(t,m)}) \leq \lambda_{\max}$ for all $1\leq m \leq M$ and $t\geq 1$. Define $\nabla{\mL}_{\max} = \max_{t,m}\|\sxy^{(t,m)} - \sxx^{(t,m)}\ols\|$. If $0<\alpha< (M\lambda_{\max})^{-1}$, then we have
\begin{equation}
\label{eq:t4}
\Big\|\Ht^{(t,M)}_{\operatorname{sf}}-\ols\Big\| \leq \Big(1-\lambda_{\min}\alpha\Big)^{tM} \Big\|\Ht^{(0)}-\ols\Big\| + 2\alpha M(\lambda_{\max}/\lambda_{\min})\nabla{\mL}_{\max}.\nonumber
\end{equation}
\eet
\noindent
The detailed proof of Theorem \ref{Th4} is given in Appendix A.7. Theorem \ref{Th4} derives a numerical error bound for the difference between the sFMGD estimator and the global estimator $\ols$, under the condition that the whole dataset is given and fixed. The upper bound can be decomposed into two terms. The first term is due to the contracted distance between the initial value $\wh\theta^{(0)}$ and $\ols$, while the second term is due to the accumulated computational error. The first term suggests that the effect of initial value diminishes at an exponential rate with a contraction factor $\big(1-\lambda_{\min}\alpha\big)^M$. As $\alpha \to 0$, we should have $\big(1-\lambda_{\min}\alpha\big)^M\to 1$. Consequently, a smaller learning rate $\alpha$ leads to a slower numerical convergence rate. Besides, a larger number of mini-batches $M$ leads to a smaller contraction factor $\big(1-\lambda_{\min}\alpha\big)^M$. Furthermore, for a fixed learning rate $\alpha$ satisfying $\alpha< \lambda^{-1}_{\max}$, we can verify that $1-\lambda_{\min}\alpha \geq 1- \lambda_{\min}/\lambda_{\max}$. Thus a smaller conditional number $\big(\lambda_{\max}/\lambda_{\min}\big)$ leads to a faster convergence rate. The second term is the accumulated computational error due to mini-batch gradient updates. We find that this term is linearly bounded by the learning rate $\alpha$, the number of mini-batches $M$, the conditional number $\lambda_{\max}/\lambda_{\min}$, and the local gradient evaluated at global minimizer $\ols$ (i.e. $\|\sxy^{(t,m)}-\sxx^{(t,m)}\ols\|$). It is notable that, once the whole dataset is given and fixed, both the quantity $\lambda_{\max}/\lambda_{\min}$ and $\max_{t,m}\big\|\sxy^{(t,m)}-\sxx^{(t,m)}\ols\big\|$ are fixed and thus bounded. Then the second term can be arbitrarily small by specifying a sufficiently small learning rate $\alpha$. Note that a smaller learning rate would increase the first term by $\big(1-\lambda_{\min}\alpha\big)^M$ but decrease the second term by $\alpha M(\lambda_{\max}/\lambda_{\min})\nabla{\mL}_{\max}$. As we can see, the learning rate $\alpha$ plays opposite roles for the two different terms. Thereafter, a constant learning rate can hardly be used to achieve the best trade-off between those two terms.

Lastly, we extend the fruitful theoretical results obtained in previous subsections to a general loss function. Define $\ell(X_i,Y_i;\theta)$ to be an arbitrary loss function evaluated at sample $i$. Let $\mL_N(\theta)=N^{-1}\sum^N_{i=1} \ell(X_i,Y_i;\theta)$ denote the global loss function. Then the global optimal estimator can be defined as $\widehat{\theta}=\operatorname{\argmin}_{\theta}\mL_N(\theta)$. For example, $\mL_N(\theta)$ could be the two-times negative log-likelihood function of a generalized linear regression model. Then $\widehat{\theta}$ should be the corresponding maximum likelihood estimator. We assume that $\widehat{\theta}$ is $\sqrt{N}$-consistent. It is then of great interest to study the theoretical properties of the FMGD-type (i.e., FMGD and sFMGD) estimators under the general loss function. Let $\Ht^{(t,m)}_{\operatorname{gf}}$ be the FMGD-type estimator obtained in the $t$-th epoch on the $m$-th mini-batch under the general loss function. Then the following updating formula for the FMGD-type algorithm is given as
\beq
\begin{split}
\label{eq:gl}
\Ht^{(t,1)}_{\operatorname{gf}}&=\Ht^{(t-1,M)}_{\operatorname{gf}}-\alpha\nabla\mL_n^{(t,1)}\Big(\Ht^{(t-1,M)}_{\operatorname{gf}}\Big),\\
\Ht^{(t,m)}_{\operatorname{gf}}&=\Ht^{(t,m-1)}_{\operatorname{gf}}-\alpha\nabla\mL_n^{(t,m)}\Big(\Ht^{(t,m-1)}_{\operatorname{gf}}\Big) \mbox{ for } 2\leq m \leq M,\nonumber
\end{split}
\eeq
where $\mL_n^{(t,m)}(\theta)=n^{-1}\sum_{i\in\mS^{(t,m)}}\ell(X_i,Y_i;\theta)$ is the loss function for the $m$-th mini-batch in the $t$-th epoch, and $\nabla\mL_n^{(t,m)}(\theta)$ is the first-order derivatives of $\mL_n^{(t,m)}(\theta)$ with respect to $\theta$. Accordingly, define $\nabla^2\mL_n^{(t,m)}(\theta)$ to be the second-order derivatives of $\mL_n^{(t,m)}(\theta)$ with respect to $\theta$. Recall that for the FMGD estimator, we should have $\mS^{(t_1,m)}=\mS^{(t_2,m)}$ for any $t_1 \neq t_2$; while for the sFMGD estimator, we usually have $\mS^{(t_1,m)}\neq \mS^{(t_2,m)}$ for any $t_1 \neq t_2$. Then the theoretical properties of the FMGD-type estimator under the general loss function are given by the following theorem.
\bet
\label{Th5}{\sc (General Loss Function)}
Assume that the whole data $\{(X_i,Y_i):1\leq i \leq N\}$ are given and fixed. Assume that there exist two positive constants $0<\lambda_{\min}\leq \lambda_{\max}<+\infty$ such that $\lambda_{\min}\leq \lambda_{p}\big\{\nabla^2\mL_n^{(t,m)}(\theta)\big\}\leq \lambda_{1}\big\{\nabla^2\mL_n^{(t,m)}(\theta)\big\} \leq \lambda_{\max}$ for all $1\leq m \leq M$, $t\geq 1$ and $\theta \in \mathbb{R}^p$. Define $\nabla{\mL}_{\max} = \max_{t,m}\| \nabla\mL_n^{(t,m)}(\widehat{\theta})\|$ as the maximum of the norm of local gradient evaluated at the global minimizer $\Ht$.  Define $\alpha_t$ is the learning rate used in the $t$-th epoch. Then we have
\begin{equation}
	\begin{split}
		\Big\|\Ht_{\operatorname{gf}}^{(t,M)} - \Ht\Big\|\leq \frac{\big\|\Ht^{(0)} - \Ht\big\|}{\exp\left\{M\lambda_{\min}\left(\sum^t_{k=1}\alpha_k\right)\right\}} + M\nabla{\mL}_{\max} \left(\frac{\lambda_{\max}}{\lambda_{\min}}\right)\sum^t_{k=1} \frac{\alpha^{2}_k}{\left(\sum^{t}_{s=k+1}\alpha_s\right)}.\nonumber
	\end{split}
\end{equation}
Furthermore, if the learning rate sequence satisfies the following three conditions:(1) $\sum^\infty_{t=1}\alpha_t = \infty$, (2) $\sum^\infty_{t=1}\alpha^2_t < \infty$, and (3) $0<\alpha_1 < \lambda^{-1}_{\max}$. Then we have $\|\Ht_{\operatorname{gf}}^{(t,M)} - \Ht\| \to 0$ as $t\to \infty$.

\eet
\noindent
The detailed proof of Theorem \ref{Th5} is given in Appendix A.8. It derives a numerical error bound for the difference between the FMGD-type estimator and the global estimator under a general loss function. We find that the results in Theorem \ref{Th5} are similar with the results in Theorem \ref{Th3}. By Theorem \ref{Th5}, the initial point, mini-batch size, learning rate, and eigenvalues of $\nabla^2\mL(\theta)$ can all affect the numerical error between the FMGD-type estimator and the global estimator. However, under the assumptions $\sum^\infty_{t=1}\alpha_t = \infty$ and $\sum^\infty_{t=1}\alpha^2_t < \infty$, the numerical error can shrink to 0 for a fixed $k$ with $t\rightarrow \infty$. It implies that, as long as $t$ is sufficiently large, the resulting FMGD-type estimator should be very close to the global estimator.

\csection{NUMERICAL STUDIES}

Extensive numerical studies are presented in this section to demonstrate the finite sample performance of the FMGD-type methods. Specifically, we aim to study their finite sample performance on various types of models. Those models are, respectively, linear regression, logistic regression, poisson regression, and a deep learning model (i.e., the convolutional neural network).

\csubsection{Linear Regression}

We start with the finite sample performance of the FMGD-type methods on linear regression models from two perspectives. First, we aim to understand how the learning rate $\alpha$ would affect the statistical efficiency of the FMGD-type estimators. Second, we try to demonstrate the finite sample performance of the proposed learning rate scheduling strategy. Specifically, consider a standard linear regression model with $N=5,000$ and $p=50$. For $1 \leq i \leq N$, the predictor $X_i$ is generated from the multivariate normal distribution with mean zero and covariance matrix $\Sigma_{xx}=(\sigma_{j_1j_2})$, where $\sigma_{j_1j_2}=0.5^{|j_1-j_2|}$ for $1 \leq j_1, j_2 \leq p$. Fix $\theta=(1,1,..,1)^{\top} \in \mR^{p}$ and $\sigma^2_{\varepsilon}=1$. Then, the response vector can be generated according to \eqref{eq:ols}. Once the data are generated, the FMGD and sFMGD estimators can be computed. We also compute the SMGD estimator for comparison, in which each mini-batch is randomly generated from the whole dataset. Finally, the OLS estimator is computed as the optimal global estimator.

\begin{figure}[h]
	\centering
	\subfloat[$\alpha=0.2$]{
		\includegraphics[width=0.49\textwidth]{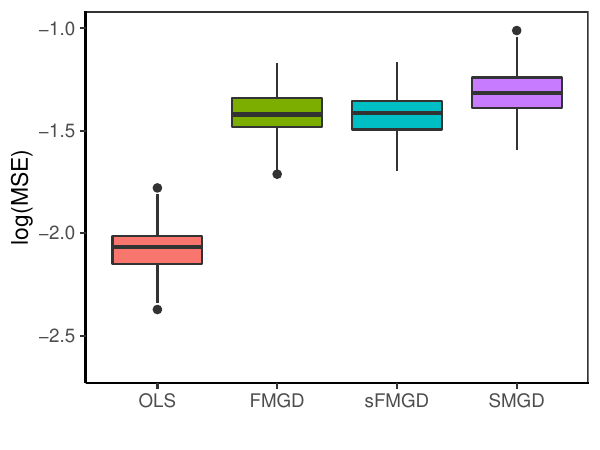}}\hfill
	\subfloat[$\alpha=0.1$]{
		\includegraphics[width=0.49\textwidth]{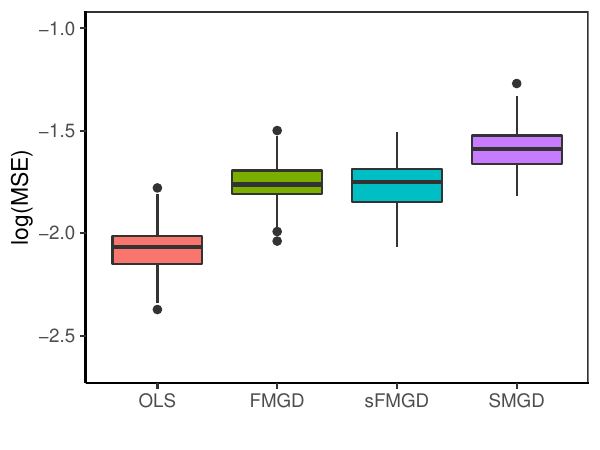}}\hfill
    \subfloat[$\alpha=0.05$]{
    		\includegraphics[width=0.49\textwidth]{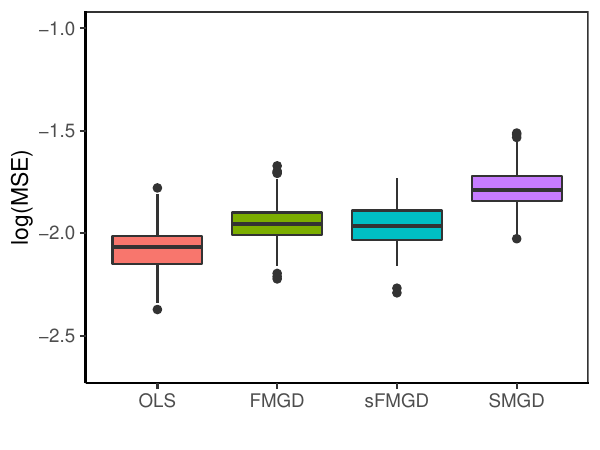}}\hfill
    \subfloat[$\alpha=0.01$]{
    		\includegraphics[width=0.49\textwidth]{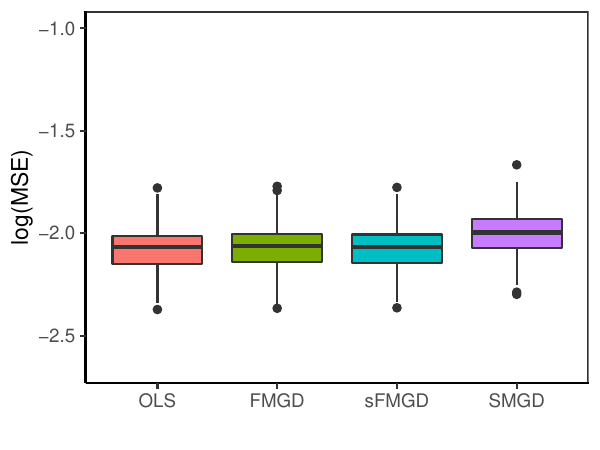}}\hfill
	\caption{The boxplots of log(MSE) values of FMGD, sFMGD, SMGD, and OLS estimators under four different fixed learning rates.}
	\label{simu_fixed}
\end{figure}

{\sc Case 1. (Constant Learning Rates)} For all the mini-batch gradient decent (MGD) estimators (i.e., FMGD, sFMGD, and SMGD), we set the mini-batch size as $n=100$ and consider four different fixed learning rates ($\alpha=0.2,0.1,0.05,0.01$). The total number of epoch iterations is fixed as $T=100$. The statistical efficiencies of the resulting estimators are then evaluated by mean squared error (MSE). For a reliable evaluation, the experiment is randomly replicated for a total of $B=200$ times for each fixed learning rate. This leads to a total of $B=200$ MSE values for each estimator, which are then log-transformed and boxplotted in Figure \ref{simu_fixed}.

By Figure \ref{simu_fixed}, we can draw the following conclusions. First, when the learning rate is not small (e.g., $\alpha=0.1$ or 0.2), the FMGD estimator demonstrates much larger MSE values than the OLS estimator. This finding is consistent with the theoretical claims in Theorem 1. That is, with a larger fixed learning rate, the FMGD estimator cannot be statistically as efficient as the whole sample estimator. Second, as the learning rate decreases, the statistical efficiency of the FMGD estimator improves, as its MSE values getting smaller and thus closer to those of the OLS estimator. These results verify our theoretical findings in Theorem \ref{Th2}. That is, the statistical efficiency of the FMGD estimator can be improved by letting $\alpha \rightarrow 0$. Next, comparing the FMGD estimator with the sFMGD estimator, we find the sFMGD estimator can achieve similar estimation performance with the FMGD estimator under the same fixed learning rate. Finally, we find both the FMGD estimator and sFMGD estimator perform quite better than the SMGD estimator. When the learning rate is small enough (i.e. $\alpha=0.01$), the performances of both the FMGD and sFMGD estimators can be very close to that of the OLS estimator.

\begin{figure}[h]
 	\centerline{\includegraphics[width=3.5in]{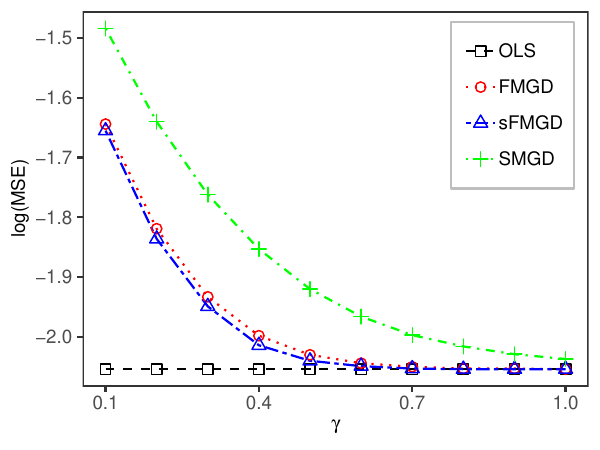}}
 	\caption{The averaged log(MSE) values of FMGD, sFMGD, SMGD, and OLS estimators using the learning rate scheduling strategy $\alpha_t=0.2t^{-\gamma}$ with $\gamma$ ranging from 0.1 to 1.}
 	\label{simu_shuffle}
 \end{figure}

 {\sc Case 2. (Learning Rate Scheduling)} We next study the proposed learning rate scheduling strategy. We fix the mini-batch size as $n=100$. Define the scheduling learning rate sequence as $\alpha_t=0.2t^{-\gamma}$ for $\gamma$ varying from 0.1 to 1. Different $\gamma$ specifications lead to different scheduling strategies. According to Theorem \ref{Th3}, we know $0.5<\gamma\leq 1$ should result in statistically efficient FMGD estimator. For each fixed $\gamma$, we evaluate this scheduling strategy for a total of $T=100$ epoch iterations. Then the statistical efficiencies of FMGD, sFMGD and SMGD estimators are evaluated in terms of MSE. The global OLS estimator is also evaluated for comparison purpose. The experiment is randomly replicated for $B = 200$ times for each fixed $\gamma$. This leads to a total of $B=200$ MSE values for each $\gamma$ specification. They are then log-transformed, averaged, and displayed in Figure \ref{simu_shuffle}.

By Figure \ref{simu_shuffle}, we can draw the following conclusions. First, for the given learning rate scheduling strategy, both FMGD and sFMGD estimators perform similarly. They both demonstrate better performances (i.e., the lower MSE values) than the SMGD estimator. Second, when the diminishing rate $\gamma$ is appropriately set (i.e., $0.5<\gamma\leq 1$), both FMGD and sFMGD estimators can achieve similar MSE values as the OLS estimator. However, if $\gamma$ is inappropriately specified (i.e., $\gamma\leq0.5$), then all MGD estimators would perform much worse than the OLS estimator. These empirical findings coordinate our theoretical claims in Theorem \ref{Th3} very well.

\begin{figure}[h]
	\centering
	\subfloat[Fixed Learning Rate]{
		\includegraphics[width=0.49\textwidth]{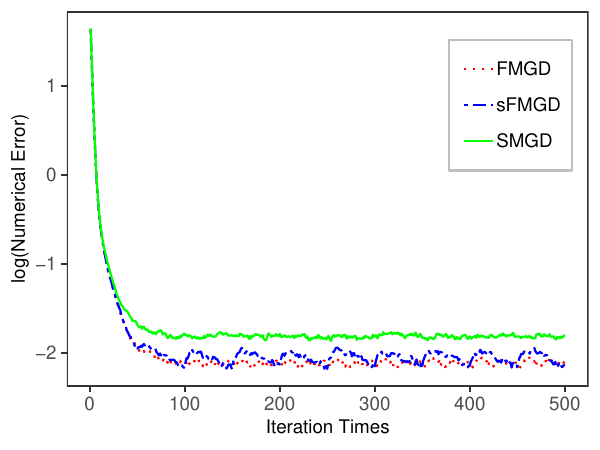}}\hfill
	\subfloat[Diminishing Learning Rate]{
		\includegraphics[width=0.49\textwidth]{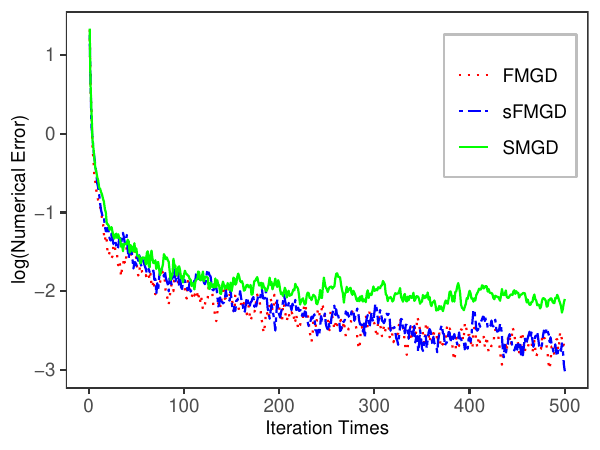}}\hfill
    \subfloat[Fixed Learning Rate]{
		\includegraphics[width=0.49\textwidth]{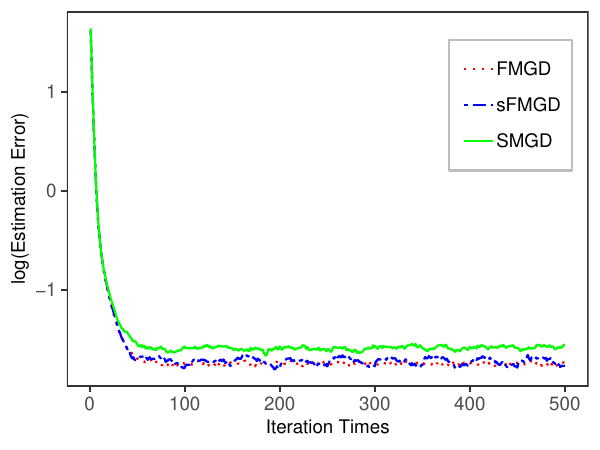}}\hfill
	\subfloat[Diminishing Learning Rate]{
		\includegraphics[width=0.49\textwidth]{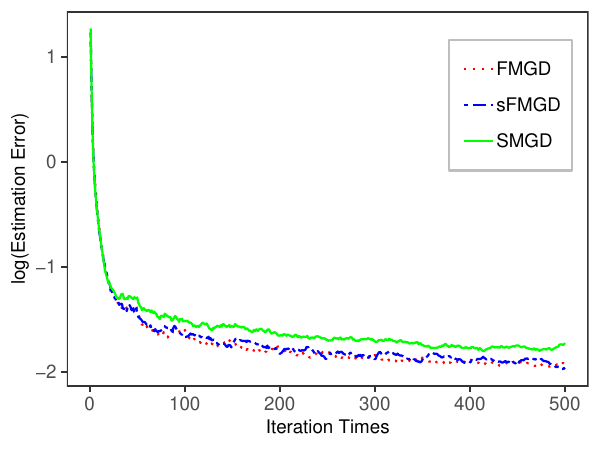}}\hfill
	\caption{The numerical and estimation errors obtained by FMGD, sFMGD, and SMGD, respectively. In the fixed learning rate example, we set $\alpha=0.1$. In the diminishing learning rate example, we set $\alpha_t=0.2t^{-\gamma}$ with $\gamma=0.6$.}
	\label{simu_conver}
\end{figure}

Lastly, we focus on the numerical convergence rates of different methods. For illustration purpose, we consider two examples: (1) the fixed learning rate with $\alpha=0.1$ and (2) the diminishing learning rate scheduling strategy $\alpha_t=0.2t^{-\gamma}$ with $\gamma=0.6$. In each example, define $\widehat{\theta}^{(t)}$ to be the estimate obtained by the MGD-type method (i.e., FMGD, sFMGD, and SMGD) in the $t$-th iteration. Define $\widehat{\theta}_{\text{ols}}$ to be the OLS estimate. We then investigate the convergence performance of different methods from two perspectives. The first perspective is the \emph{numerical error}, in which we compare the iteration-wise estimate with the global OLS estimate by $\|\widehat{\theta}^{(t)}-\widehat{\theta}_{\text{ols}}\|$. The second perspective is the \emph{estimation error}, in which we compare the iteration-wise estimate with the true parameter by $\|\widehat{\theta}^{(t)}-\theta\|$. The experiment is randomly repeated for a total of $B=200$ times, and the averaged numerical errors and estimation errors over 200 replications can be computed. Detailed results are given in Figure \ref{simu_conver} in log-scale. We find both FMGD and sFMGD perform much better than SMGD in both the numerical error and estimation error.

\newpage
\csubsection{General Loss Functions}

We next demonstrate the finite sample performances of FMGD-type algorithms on various general loss functions. The whole experiments are designed similarly as those in the Section 4.1. The only difference is that, the loss function is replaced by more general ones. Specifically, we use the negative two-times log-likelihood function as the general loss function for the following two examples.

\begin{enumerate}[{[1]}]
\item {\sc (Logistic Regression)} Logistic regression has been widely used to model the binary response. In this example, we consider the whole sample size $N=5,000$ and the dimension of predictors $p=50$. We first follow Section 4.1 to generate the predictor $X_i$ with $1\leq i \leq N$. Then fix $\theta=(0.1,0.1,...,0.1)^{\top}\in \mR^{p}$. The response $Y_i$ is then generated from a Bernoulli distribution with the probability given as $P(Y_i=1|X_i,\theta)=\exp(X_i^{\top}\theta)/\{1+\exp(X_i^{\top}\theta)\}$.
\item {\sc (Poisson Regression)} Poisson regression is often used to deal with the count responses. In this example, we also assume $N=5,000$ and $p=50$. The predictor $X_i$ is generated similarly as in Section 4.1. Let $\theta=(0.02,0.02,...,0.02)^{\top}\in \mR^{p}$. Then, the response $Y_i$ is generated from a Poisson distribution given as $P(Y_i|X_i,\theta)=\lambda_i^{Y_i}\exp(\lambda_i)/Y_i!$, where $\lambda_i=\exp(X_i^{\top}\theta)$.
\end{enumerate}

\begin{figure}[h]
	\centering
	\subfloat[Logistic Regression]{
		\includegraphics[width=0.46\textwidth]{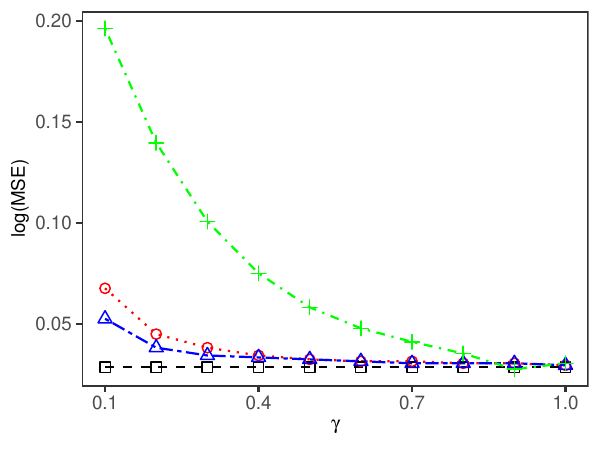}}\hfill
	\subfloat[Poisson Regression]{
		\includegraphics[width=0.46\textwidth]{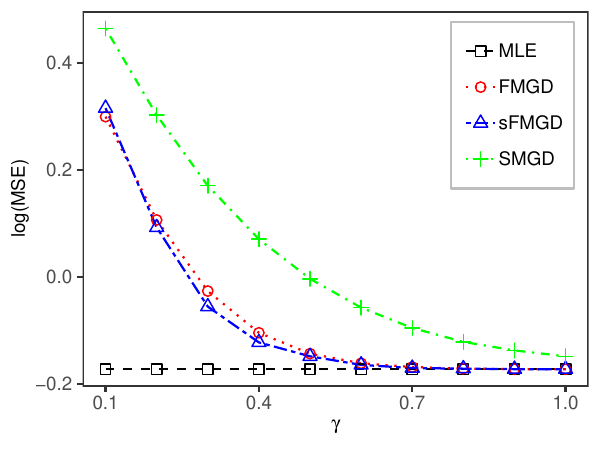}}\hfill
	\caption{The averaged log(MSE) values of FMGD, sFMGD, SMGD, and MLE estimators under the logistic regression model and poisson regression model. For the MGD methods, the diminishing learning rate scheduling strategy $\alpha_t=0.2t^{-\gamma}$ is used with $\gamma$ ranging from 0.1 to 1.}
	\label{simu_general}
\end{figure}

Once the data are generated, various MGD estimators (i.e., FMGD, sFMGD, and SMGD) are computed. The whole sample maximum likelihood estimator (MLE) is also computed for comparison purpose. For the MGD methods, various learning rate scheduling strategies are applied. The experiments for the MGD methods are conducted in the similar way as in Section 4.1. The detailed results are given in Figure \ref{simu_general}. We find the conclusions are quantitatively similar to those in Section 4.1. Specifically, for both the logistic regression and poisson regression models, the FMGD and sFMGD estimators show clear advantages over the SMGD estimator. Moreover, when the diminishing rate $\gamma$ is appropriately chosen, the FMGD and sFMGD estimators could be statistically as efficient as the whole sample optimal estimator (i.e., the MLE). The outstanding numerical performance of the sFMGD estimator is expected, because it is extremely similar with the FMGD estimator. The only difference is that, the shuffling operation should be conducted for every epoch for sFMGD. The outstanding performances of FMGD and sFMGD estimators are also consistent with the theoretical claims in Theorem \ref{Th5}.

\csubsection{Sampling Cost}

In this subsection, we focus on the time cost of FMGD and sFMGD methods. As demonstrated in the previous subsection, both methods are very competitive in terms of statistical efficiency. In particular, with appropriately designed learning rate scheduling strategy, both the FMGD and sFMGD estimators can be statistically as efficient as the global estimator. Then it is of great interest to query about the difference between these two estimators. In fact, we are going to demonstrate in this subsection that there indeed exists a critical difference, i.e., the time cost.

Note that the FMGD methods (both FMGD and sFMGD) are designed for datasets of massive sizes. In this case, the data have to be placed on the hard drive, instead of the memory. If the sFMGD method is applied, then the sample indices in each mini-batch have to be randomly shuffled for each epoch iteration. As a consequence, the samples in each mini-batch have to be obtained from random positions on the hard drive from epoch to epoch. This leads to a significant time cost for hard drive addressing. In contrast, if the FMGD method is applied, then the whole sample only needs to be shuffled once. Then, according to the pre-specified mini-batch size, the whole sample can be split sequentially into a sequence of $M$ non-overlapping mini-batches. Each mini-batch is then formed into one packaged data file. By doing so, we should pay two prices. The first price is the time cost for data packaging. Our numerical experiments suggest that, this is a time cost no more than that of two epochs for a standard sFMGD updating. Therefore, the time cost due to data packaging can be very negligible. The second price is the storage cost on the hard drive. By data packaging, we effectively duplicate the data size on the hard drive. However, given the outstanding storage capacity of modern hard drives, we find this cost is practically very acceptable. Once the mini-batches are packaged, they would be sequentially placed on the hard drive. Subsequently, they can be read into the memory at a much faster speed than those used in the sFMGD method. To demonstrate the computational advantage of the FMGD method over the sFMGD method, we conduct the following experiments.

\begin{figure}[h]
	\centerline{\includegraphics[width=3.5in]{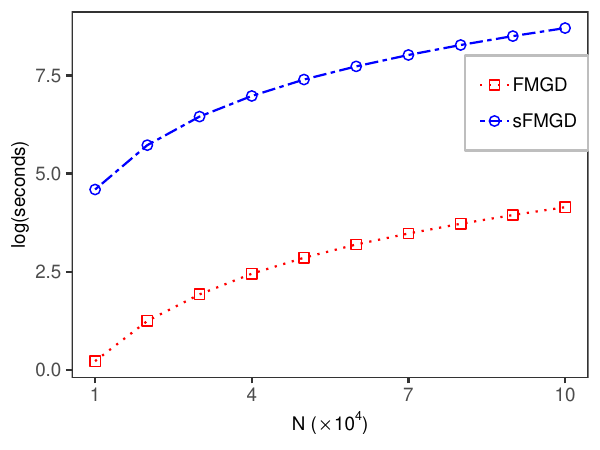}}
	\caption{The time costs (in log-scale) per epoch consumed by FMGD and sFMGD methods for different sample sizes $N$.}
	\label{simu_time}
\end{figure}

We generate the dataset similarly as in Section 4.1, but with a larger sample size $N$. Specifically, we consider $N=\kappa10^4$ with $\kappa$ varying from 1 to 10. The generated data are then placed on the hard drive in CSV format. Fix the mini-batch size as $n=100$. We then compute the FMGD and sFMGD estimators for a linear regression model in the similar way as in Section 4.1. The total time costs consumed by the FMGD and sFMGD methods under different sample sizes are averaged over $B=20$ random replications. Then, the averaged time costs in each epoch are plotted in Figure \ref{simu_time} in log-scale. As shown in Figure \ref{simu_time}, the time costs consumed by the FMGD method are much smaller than those consumed by the sFMGD method. In addition, as the whole sample size increases, the time costs consumed by the FMGD and sFMGD methods both increase. However, the sFMGD method increases at a much faster speed than the FMGD method (recall the difference of time cost reported in Figure \ref{simu_time} is in log-scale). These empirical findings confirm that the FMGD method is computational more efficient than the sFMGD method.


\csubsection{Deep Neural Networks}\label{DL}

\begin{figure}[h]
	\centerline{\includegraphics[width=6in]{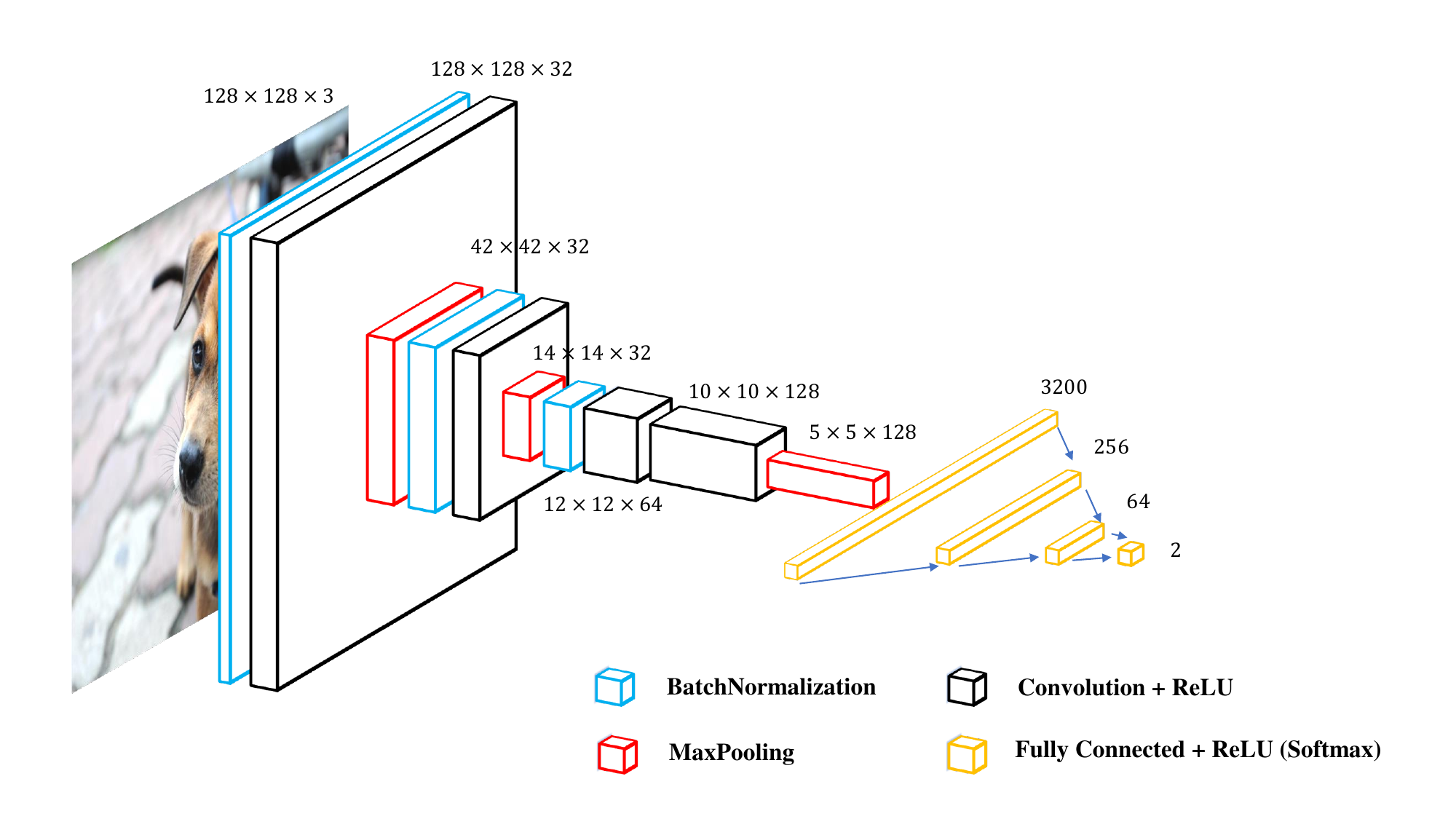}}
	\caption{The structure of designed CNN model.}
	\label{cnn}
\end{figure}

We next demonstrate the performance of the FMGD-type methods on a deep neural network model. Compared with the models studied in the previous subsections, deep neural networks are considerably more sophisticated. Specifically, we consider a convolutional neural network (CNN) for image data analysis. We study in this example the famous CatDog dataset, which was first released by Kaggle in 2013. The dataset contains 15,000 training images and 10,000 validation images. Those images belong to two classes (i.e., cats and dogs). The objective here is automatic image classification (cats v.s. dogs). To solve this problem, a simple convolutional neural network is designed in Figure \ref{cnn}.

This is a standard CNN model. Specifically, the input is a fixed-size $128\times128$ RGB image. Subsequently, the image (i.e. $128\times128\times 3$ tensor) is passed through three batch-normalization-convolution-maxpooling blocks. In the first block, we utilize $11\times 11$ convolution filters with the same padding for preliminary feature extraction. We then down sample the features with maxpooling of stride 3. In the second block, we change the size of convolution filters to $5\times 5$ and keep the stride equal to 3 in the maxpooling layer. In the third block, we set two consecutive convolutional layers with $3\times 3$ convolution filters of amount $64$ and $128$ respectively. They are then followed by a maxpooling layer with stride 2. Finally, these feature extraction blocks are followed by three fully-connected layers, which have 256, 64 and 2 channels, respectively. The last fully-connected layer is also combined with a softmax activation function, which links the feature map to the output probability. This leads to a CNN model with 16 layers and a total of 965,800 parameters.

\begin{figure}[h]
	\centerline{\includegraphics[width=6in]{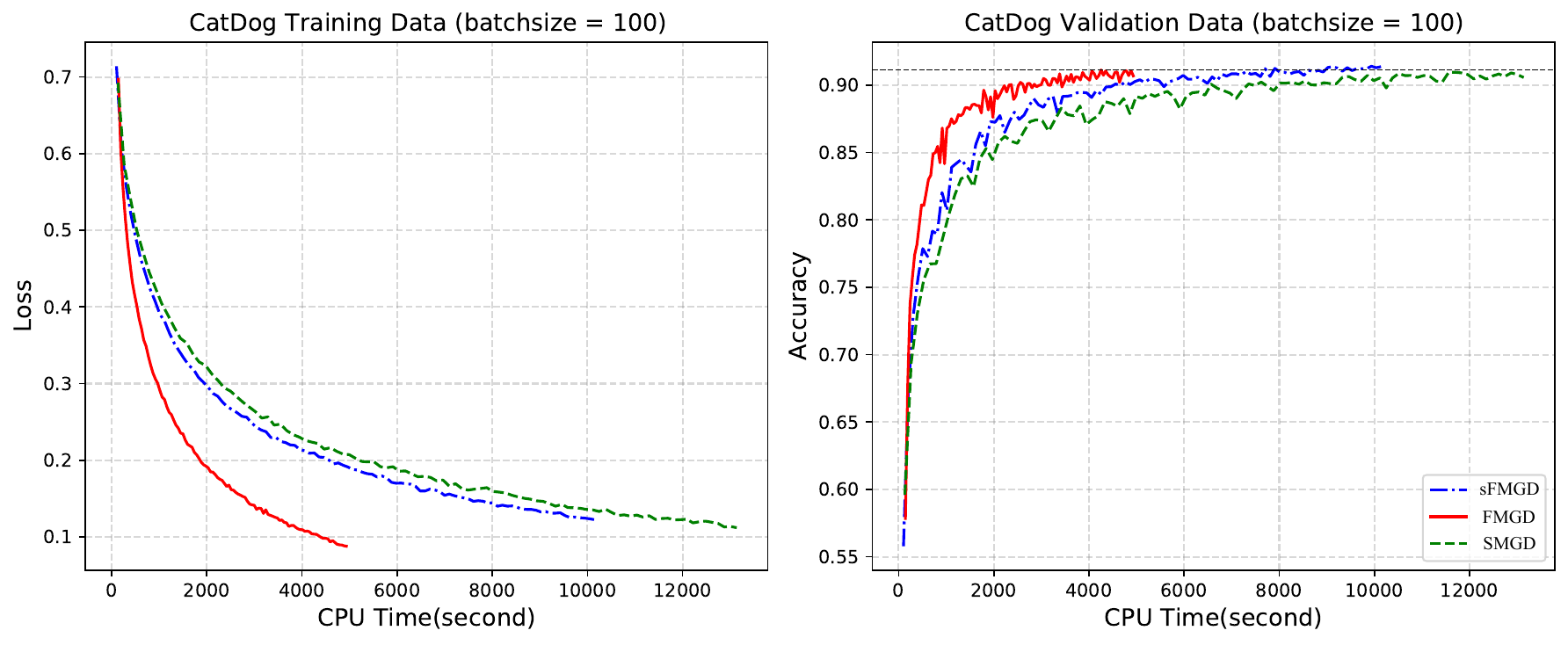}}
	\caption{The left panel shows the averaged training loss curves of FMGD, sFMGD and SMGD algorithms. The right panel shows the averaged validation accuracy curves of these three MGD algorithms. The horizontal black dashed line denotes the best validation accuracy that can be achieved by the FMGD algorithm.}
	\label{simu_catdog(time)}
\end{figure}

Various MGD algorithms (i.e., FMGD, sFMGD, and SMGD) are used to train this model. For each MGD algorithm, we set the mini-batch size as $n = 100$ and run a total of $T = 100$ epochs. The resulting training losses and validation accuracies are monitored and averaged over $B=20$ randomly replicated experiments. The averaged training loss curves and validation accuracy curves are given in Figure \ref{simu_catdog(time)}. By the left panel in Figure \ref{simu_catdog(time)}, we find that the FMGD method demonstrates the fastest performance in loss reduction. As a consequence, it achieves the fastest accuracy improvement in the validation dataset; see the right panel of Figure \ref{simu_catdog(time)}. Comparatively, the loss reduction speed of sFMGD is slower than that of the FMGD. Consequently, its prediction accuracy does not improve as fast as FMGD. Nevertheless, the final prediction accuracy of sFMGD is slightly better than that of FMGD (i.e., 91.41\% for sFMGD v.s. 91.13\% for FMGD). The performance of SMGD is the worst from all perspectives.

\csection{CONCLUDING REMARKS}

We study in this work a fixed mini-batch gradient decent (FMGD) algorithm. The key difference between FMGD and SMGD is how the mini-batches are generated. For FMGD, the whole sample is split into multiple non-overlapping partitions (i.e., mini-batches). Once the mini-batches are formed, they are fixed and then repeatedly used throughout the rest of the algorithm. Consequently, the mini-batches used by FMGD are dependent with each other. However, for SMGD the mini-batches are independent with each other either marginally or conditionally. This difference makes FMGD enjoy faster numerical convergence and better statistical efficiency.

To conclude this article, we discuss here a number of interesting directions for future research. First, we study the theoretical properties of the FMGD-type estimators under a fixed $p$ setup. Note that modern statistical analyses often encounter problems with ultrahigh dimensional features. It is then of great interest to investigate the theoretical properties of the FMGD-type estimators with diverging $p$. Second, many FMGD variants (e.g., momentum, Adagrad, RMSprop, Adam) have been popularly used in practice. However, it seems to us that, their theoretical properties under the fixed mini-batch setup remain unknown. This should also be another interesting topic for future study.

\section*{Acknowledgement}
The authors thank the Editor, Associate Editor, and two anonymous reviewers for their careful reading and constructive comments. This work is supported by National Natural Science Foundation of China (No.72001205), the Fundamental Research Funds for the Central Universities and the Research Funds of Renmin University of China (21XNA026).

\section*{Conflict of Interest}
The authors report there are no competing interests to declare.

\newpage
\bibliographystyle{asa}
\bibliography{ref}
\end{document}